\documentclass[aps,preprint,amsmath,amssymb,nofootinbib]{revtex4}

\usepackage{hyperref}
\usepackage{graphicx}
\usepackage{slashed}

\begin{document}

\title{Democratic Neutrino Theory}


\author{Dmitry Zhuridov}%
\email{dmitry.zhuridov@wayne.edu}%
\email{jouridov@mail.ru}%
\affiliation{Department of Physics and Astronomy, Wayne State University,
Detroit, MI 48201}

\date{\today} 

\begin{abstract}

New theory of neutrino masses and mixing is introduced. This theory is based on a simple $S_3$ symmetric  democratic neutrino mass matrix, and predicts the neutrino mass spectrum of normal ordering.  
Taking into account the matter effect and proper averaging of the oscillations, this theory agrees with the variety of atmospheric, solar and accelerator neutrino data. Moreover, the absolute scale of the neutrino masses $m\approx0.03$~eV is determined in this theory, using the atmospheric neutrino oscillation data. 
In case of tiny perturbations in the democratic mass matrix only one this scale parameter $m$ allows to explain the mentioned above neutrino results, and the theory has huge predictive power. 
\end{abstract}

\maketitle

\section{Introduction}

The neutrinos were proposed by W. Pauli in 1930~\cite{Pauli:2000ak} and first detected by C.~
Cowan and F.~
Reines in 1956~\cite{Cowan:1992xc}. 
Later the solar neutrino deficit~\cite{Davis:1968cp}, the atmospheric neutrino oscillations~\cite{Fukuda:1998mi}, and other neutrino oscillation results~\cite{PDG2012} have revealed that the neutrino masses are nonzero. 
However the questions about the  absolute scale of their masses, their type (whether Dirac or Majorana), etc., still remain unanswered.

Besides the reactor~\cite{Mention:2011rk}, Gallium~\cite{Giunti:2012tn,Abdurashitov:2005tb} and few other neutrino anomalies, the available neutrino data can be explained within the phenomenological model of three light neutrinos with two mass splittings (squared mass differences) of close scales: $\Delta m_s^2\sim10^{-4}$~eV$^2$ and $\Delta m_a^2\sim10^{-3}$~eV$^2$~\cite{PDG2012}. We call it the conventional neutrino theory (CNT). In case of Majorana (Dirac) neutrinos the masses and mixings are generically parametrized in CNT by 9 (7) free parameters: 3 masses, 3 mixing angles and 3 phases (1 phase). 

In this paper we show that the variety of the neutrino data can be explained in a model with three Majorana neutrinos with simple $S_3$ symmetric (in the leading order) ``democratic" mass matrix. This democratic neutrino theory (DNT), which was briefly presented in Ref.~\cite{Zhuridov:2013ika}, predicts the normal hierarchy of the neutrino mass spectrum, and allows to determine from the atmospheric neutrino data the absolute scale of the neutrino masses of $m\approx0.03$~eV. 

In the case of slightly perturbed democratic mass matrix, DNT has huge predictive power, and only one its parameter $m$ is enough to  determine to a good precision the available oscillation data. However there is a single large mass splitting. If it governs the atmospheric neutrino oscillations than the second mass splitting is much less than $10^{-4}$~eV$^2$.  For this reason we give up the interpretation of the data obtained by the KamLAND experiment~\cite{Araki:2004mb,Gando:2010aa,Gando:2013nba} in terms of the neutrino oscillations due to their mass splitting of order $10^{-4}$~eV$^2$,%
\footnote{We take an advantage of absence of another independent experiment similar to KamLAND, which can verify this interpretation.} 
and provide a nonstandard explanation of the solar neutrino data. 
Also in the discussed case the neutrino mass matrix is essentially different from the tri-bimaximal one~\cite{Wolfenstein:1978uw,Fritzsch:1995dj,Harrison:2002er,Harrison:2002kp,Xing:2002sw}, which motivates us to use alternative explanation of several atmospheric and accelerator neutrino observables. 

On the other hand, for relatively large deviation from $S_3$ symmetric mass matrix a standard interpretation of the neutrino data may come into play, while the predictive power is reduced. 

In the next section DNT is introduced, and the neutrino masses and mixing are derived.  
Further we discuss the neutrino oscillations in section~\ref{Neutrino oscillations} and alternative explanations of the neutrino experimental data in the framework of DNT in section~\ref{Neutrino experiments}. Then we compare DNT with CNT in section~\ref{Discussion}, discuss the predictions of DNT in section~\ref{Predictions}, and conclude.

\section{Democratic Masses and Mixing}

A lowest order democratic quark mass matrix proportional to
\begin{eqnarray}
	\left(  
\begin{array}{ccc}
	1 & 1 & 1 \\
		1 & 1 & 1 \\
			1 & 1 & 1 

\end{array}
		\right)
\end{eqnarray}
was discussed in the context of $S_{3}(L)\times S_{3}(R)$ symmetry of massless quarks in 
the $SU(2)_L\times SU(2)_R\times U(1)$ gauge model in Ref.~\cite{Harari:1978yi}. 
Possible relation of this matrix to the leptonic mixing was investigated in Ref.~\cite{Fritzsch:1995dj}. 
The two $S_3(L)$ invariant mass matrixes for Majorana neutrinos 
\begin{eqnarray}\label{eq:2mass_matrices}
\left( \begin{array}{ccc}
    1 & 0 & 0 \\ 
    0 & 1 & 0 \\ 
    0 & 0 & 1 \\ 
  \end{array} \right), 
  \qquad 
	\left( \begin{array}{ccc}
    0 & 1 & 1 \\ 
    1 & 0 & 1 \\ 
    1 & 1 & 0 \\ 
  \end{array} \right)	
\end{eqnarray}
were discussed in Ref.~\cite{Fukugita:1998vn} with focus on the first of them. 
The second of the matrices in Eq.~\eqref{eq:2mass_matrices} was considered in Ref.~\cite{Nicolaidis:2013hxa} in the context of brane models.  
%
In DNT we consider this matrix as the leading order neutrino mass matrix in usual low-energy $SU(3)_{color}\times U(1)_{em}$ theory. Then we  add either sizable (of order few \%) or tiny (orders of magnitude less than 1\%) perturbations. These two cases require different explanation of the neutrino data, see the following.

\subsection{DNT to the leading order}

Consider the mass term for three Majorana neutrinos
\begin{eqnarray}\label{eq:Lm}
	\mathcal{L}_m^\nu	=	-\frac{1}{2}\,	\sum_{\alpha\beta}\bar\nu_{\alpha L}^c M_{\alpha\beta}\nu_{\beta L} + {\rm H.c.},
\end{eqnarray}
where $\alpha,\beta=e,\mu,\tau$ are the flavor indices, $c$ denotes charge conjugation, and 
\begin{eqnarray}\label{eq:mass_matrix_general}
	M	=	 M_0 + o(m)
\end{eqnarray}
is the neutrino mass matrix in the flavor basis, where
\begin{eqnarray}\label{eq:mass_matrices}
	M_0	=	 m\left( \begin{array}{ccc}
    0 & 1 & 1 \\ 
    1 & 0 & 1 \\ 
    1 & 1 & 0 \\ 
  \end{array} \right)
\end{eqnarray}
is a {\it democratic} mass matrix,\footnote{This matrix may originate from the possible intrinsic structure of leptons, which we discuss elsewhere. Also this generic framework may help to determine the small deviations in  Eq.~\eqref{eq:mass_matrix_general}.} 
which is invariant under the permutation group of three elements $S_3$~\cite{Harari:1978yi,Fritzsch:1995dj,Fukugita:1998vn,Fujii:2002jw,Mohapatra:1998rq,Harrison:2003aw,Nicolaidis:2013hxa}, and is defined by the only mass scale parameter $m$. Possible perturbations to this leading order mass matrix are included in the little-o term $o(m)$ in Eq.~\eqref{eq:mass_matrix_general}, and they are discussed below.  
The eigenvectors of $M_0$ form the neutrino mixing matrix, which can be written as\footnote{Note that Eq.~\eqref{eq:U} is different from the ``democratic mixing pattern" discussed in literature, see Ref.~\cite{Garg:2013xwa} and references therein.}
\begin{eqnarray}\label{eq:U}
	U_0	=	\left(	  \begin{array}{ccc}
     \frac{1}{\sqrt{2}} & \frac{1}{\sqrt{6}} & \frac{1}{\sqrt{3}} \\ 
    -\frac{1}{\sqrt{2}} & \frac{1}{\sqrt{6}} &  \frac{1}{\sqrt{3}}  \\ 
    0 & -\frac{2}{\sqrt{6}} & \frac{1}{\sqrt{3}} \\ 
  \end{array}	\right),	
\end{eqnarray} 
and makes the transformation from flavor to mass basis as
\begin{eqnarray}
	\nu_{\alpha L}=	\sum_{i=1,2,3} (U_0)_{\alpha i}\nu_{i L},		
\end{eqnarray}
where the subindex $0$ refers to the leading order approximation. 
The last column of $U_0$ corresponds to the larger eigenvalue of $M_0$, and the first row does not include zero to explain the solar neutrino data, see section~\ref{Solar neutrinos}. It is convenient to parametrize Eq.~\eqref{eq:U} using only two angles as
\begin{eqnarray}\label{eq:Uparametr}
	U_0	=	R_{12}(\tilde\theta_{12}) \otimes R_{23}(\tilde\theta_{23})	=	 
	\left(   \begin{array}{ccc}
    c_{12} & s_{12}c_{23} & s_{12}s_{23} \\ 
    -s_{12} & c_{12}c_{23} & c_{12}s_{23} \\ 
    0 & -s_{23} & c_{23} \\ 
  \end{array}	\right)
\end{eqnarray} 
where $c_{ij}\equiv\cos\tilde\theta_{ij}$, $s_{ij}\equiv\sin\tilde\theta_{ij}$, $\tilde\theta_{12}=45^\circ$, $\tilde\theta_{23}=\pi/2-\arctan(1/\sqrt{2})\approx54.7^\circ$, and tilde denotes our parametrization. 
Notice that in Eq.~\eqref{eq:Uparametr} we used the opposite order of multiplication of the Euler rotation matrixes $R_{ij}$ with respect to the ordinary tri-bimaximal matrix~\cite{Wolfenstein:1978uw,Fritzsch:1995dj,Harrison:2002er,Harrison:2002kp,Xing:2002sw}.

In the case of standard parametrization of the neutrino mixing matrix \cite{PDG2012,Kayser:2002qs}, Eq.~\eqref{eq:U} is reproduced by 
\begin{eqnarray}\label{eq:Ustandard} 
	U_\text{standard}	=	R_{23}(\theta_{23})  \otimes   R_{13}(\theta_{13})  \otimes   R_{12}(\theta_{12})	
\end{eqnarray} 
with  $\theta_{12}=30^\circ$, $\theta_{23}=45^\circ$, $\theta_{13}=\arcsin(1/\sqrt{3})\approx35.3^\circ$, and no $CP$ violating phases.

By diagonalizing the mass matrix $M_0$ with the matrix $U_0$ and expressing $\nu_{iL}$ through the Majorana spinors $\psi_i=\psi_i^c$ as
\begin{eqnarray}
	\nu_{iL}	\equiv	\frac{1}{2}  (1-\gamma_5)	\psi_i,
\end{eqnarray}
Eq.~\eqref{eq:Lm} can be rewritten as
\begin{eqnarray}\label{eq:Lmass}
	\mathcal{L}_m^\nu	=	
	-\frac{1}{2}\,	\sum_i	s_i m_i  \bar\psi_i \psi_i,
\end{eqnarray}
where $m_1=m_2=m_3/2\equiv m$, and $s_1=s_2=-s_3=-1$ are the sign factors, which can be absorbed by the transformation $\psi_j \to i\gamma_5\psi_j^\prime$ ($j=1,2$). On relation of the sign factors to the neutrino $CP$ properties and mixing matrix see section 2.3.2 in Ref.~\cite{Doi:1985dx} and references therein. The resulting neutrino mass spectrum
\begin{eqnarray}\label{eq:spectrum}
	\{ m, m, 2m \}
\end{eqnarray}
has normal ordering and two degenerate values, which results in the only mass splitting 
\begin{eqnarray}\label{eq:Deltam2}
	\Delta m^2 = 3 m^2.
\end{eqnarray}

\subsection{Tiny perturbations}

Similarly to the great variety of broken symmetries in nature, it is naturally that the degeneracy in Eq.~\eqref{eq:spectrum} is violated by a small perturbations in Eq.~\eqref{eq:mass_matrix_general}.  
Suppose that these perturbations are {\it tiny}. This leads to a spectrum
\begin{eqnarray}\label{eq:spectrum_perturbed}
	\{ m, m+\delta m, 2m \},
\end{eqnarray}
where $0<\delta m/m$ measures the violation of ``full democracy" (in section~\ref{Solar neutrinos} we consider $\delta m/m\ll10^{-4}$ for effective suppression of the respective oscillations in the Sun), and we ignore possible deviation in the eigenvalue $2m$ since this does not significantly effect the large neutrino mass splitting $\Delta m^2$. As a result, for the two splittings we have: 
\begin{eqnarray}
	\Delta\mu^2	\equiv	(m+\delta m)^2 - m^2	\approx2m\delta m
\end{eqnarray}
and $\Delta m^2 = 3m^2$. To consider the neutrino oscillations with reasonable accuracy the neutrino mixing matrix can be written as
\begin{eqnarray}\label{eq:Uperturbed}
U	=	\left(	  \begin{array}{ccc}
     \frac{1}{\sqrt{2}} & \frac{1}{\sqrt{6}} & \frac{1}{\sqrt{3}} \\ 
    -\frac{1}{\sqrt{2}} & \frac{1}{\sqrt{6}} &  \frac{1}{\sqrt{3}}  \\ 
    \lambda\left(\frac{\delta m}{m}\right)^{1/2} & -\frac{2}{\sqrt{6}} & \frac{1}{\sqrt{3}} \\ 
  \end{array}	\right),
\end{eqnarray} 
where $\lambda$ is a constant of order one, and the deviations in nonzero elements are neglected since all these elements are large. 

In section~\ref{Neutrino experiments} we show how atmospheric, solar and accelerator neutrino observables can be explained using only parameter $m$, neglecting small effects of other parameters.

\subsection{Sizable perturbations}

Consider the case of {\it sizable} perturbations in the mass matrix $M$, which lead to the spectrum
\begin{eqnarray}
	\{ m+x_1, m+x_2, 2m+x_3 \},
\end{eqnarray}
where the deviations $x_i$ ($x_2>x_1$) can be written as
\begin{eqnarray}\label{eq:x_i}
	x_i = \xi_i \sqrt{\Delta m_s^2}
\end{eqnarray}
with the small, but essentially larger than $\delta m$ in Eq.~\eqref{eq:spectrum_perturbed}, mass scale $\sqrt{\Delta m_s^2}$ ($0<\sqrt{\Delta m_s^2}\ll m$), and the dimensionless parameters $\xi_i$ of order unity ($\xi_2-\xi_1\ll1$). Then the mass splitting 
\begin{eqnarray}\label{eq:Delta_a^2}
	\Delta m_a^2 \equiv	(2m+x_3)^2 - (m+x_1)^2
\end{eqnarray}
can be rewritten as
\begin{eqnarray}
	\Delta m_a^2 	=	3m^2	+	\Delta m_s^2 (\xi_1^2-\xi_3^2) 	+	{\cal O} \left(\frac{\Delta m_s^4}{\Delta m_a^2}\right),
\end{eqnarray}
and we end up with the spectrum 
\begin{eqnarray}
	\left\{ m_1, \sqrt{m_1^2+\Delta m_s^2}, \sqrt{m_1^2+\Delta m_a^2} \right\},
\end{eqnarray}
where $m_1\equiv m + x_1\approx \sqrt{\Delta m_a^2/3}$. The two splittings $\Delta m_s^2$ and $\Delta m_a^2$ may correspond to the ordinary solar and atmospheric neutrino mass splittings, respectively. (That is why we used this notation for them in Eqs.~\eqref{eq:x_i} and \eqref{eq:Delta_a^2}). 
The neutrino mixing matrix has sizable corrections in relation to Eq.~\eqref{eq:U}. In particular, the $U_{e3}$ element may be significantly smaller, and there may be essential $CP$ violation. 
Thereby, ordinary interpretation of the neutrino experimental data may be applied. However in the following sections we will concentrate on more interesting and predictive case of tiny perturbations, which requires an alternative explanation of the neutrino data.

In both cases of either tiny or sizable perturbations the neutrino mass spectrum has normal ordering, and the absolute neutrino mass scale is about $\sqrt{\Delta m^2/3}$. 
Using the atmospheric neutrino mass splitting $\Delta m_a^2=(2.06-2.67)\times10^{-3}$~eV$^2$ (at 99.73\% CL)
~\cite{PDG2012}\footnote{We consider this result as approximate since it was derived in a completely different neutrino mass and mixing scheme.}, the absolute neutrino mass scale can be determined as
\begin{eqnarray}\label{eq:nu_mass_scale}
	m\approx0.03~\text{eV}.
\end{eqnarray}

\section{Neutrino oscillations}\label{Neutrino oscillations}

The 
probabilities of the flavor neutrino oscillations in vacuum can be written as
\begin{eqnarray}\label{eq:Pnu}
	&&P_{\nu_\alpha\to\nu_\beta}(L,E)	=	\sum_i  |U_{\alpha i}|^2 |U_{\beta i}|^2	\nonumber\\
	&&+	2\sum_{i>j}  |U_{\alpha i}^*U_{\beta i} U_{\alpha j}U_{\beta j}^*|	\cos\left(2\pi\frac{L}{L_{ij}^\text{osc}} 
	-\phi_{\beta\alpha;ij}\right),
\end{eqnarray}
where $L_{ij}^\text{osc}  =  4\pi  E/\Delta m_{ij}^2$ is the oscillation length, and $\phi_{\beta\alpha;ij}	=	\arg(U_{\alpha i}^*U_{\beta i} U_{\alpha j}U_{\beta j}^*)$ is the constant phase. 
Using the leading order democratic neutrino mass spectrum in Eq.~\eqref{eq:spectrum} and mixing in Eq.~\eqref{eq:U} we have the ``democratic" oscillation probabilities of the same size
\begin{eqnarray}\label{eq:PoscModel}
	P_{\nu_\alpha\to\nu_\beta}(L,E)
							=	\frac{4}{9}		\sin^2 \left( 	\frac{\Delta m^2 L}{ 4E}	\right), 	
\end{eqnarray}
where $\beta\neq\alpha$. 


In case of the spectrum of Eq.~\eqref{eq:spectrum_perturbed} the probabilities $P_{\nu_\alpha\to\nu_\beta}$ are slightly different from Eq.~\eqref{eq:PoscModel} due to the oscillation terms with $\Delta\mu^2$, while the results in Eq.~\eqref{eq:PoscModel} are reproduced in the limit $\Delta \mu^2 L/4E\to0$. In particular, for the electron neutrino survival probability we have
\begin{eqnarray}\label{eq:P_surviv}
	P_{\nu_e\to\nu_e}^\text{survival}
							=	\frac{7}{18}	+	\frac{4}{9}	 \cos \left( 	\frac{\Delta m^2 L}{ 2E}	\right)	+	\frac{1}{6}  \cos \left( 	\frac{\Delta \mu^2 L}{ 2E}	\right). 	
\end{eqnarray}

\section{Neutrino experiments}\label{Neutrino experiments}

In this section we discuss the explanation within DNT (in its case with tiny perturbations) of the data obtained by the neutrino experiments of several types.

\subsection{Atmospheric neutrinos} \label{Atmospheric neutrinos}

The atmospheric neutrino flux dominates by the neutrinos with the energies in the range $0.1-10^2$ GeV \cite{Giunti:2007ry}.  
Correspondingly, for the oscillations due to $\Delta m^2\approx 3m^2$ with $m$ from Eq.~\eqref{eq:nu_mass_scale} the oscillation length
is in the range $10^2-10^5$~km, which is relevant for the terrestrial experiments. However the oscillations due to $\Delta \mu^2 \ll 10^{-6}$~eV$^2$ can not be observed since their oscillation length $L_{12}^\text{osc}\gg10^5$~km significantly exceeds the Earth diameter. For this reason, in DNT with tiny perturbations the atmospheric neutrino oscillation probabilities are well described by Eq.~\eqref{eq:PoscModel}. 
Consider how this equation may explain 
the significant zenith-angle deficit of $\nu_\mu$ observed by the Super-Kamoikande (SK) detector~\cite{Ashie:2005ik,Ashie:2004mr,Abe:2006fu}. We remark that in CNT this deficit was interpreted as the 2-flavor $\nu_\mu \to \nu_\tau$ oscillations, taking into account the suppression of $\nu_e \to \nu_{\mu,\tau}$ oscillations by a small $\theta_{13}$. 

In DNT this difference between the $e$-like and $\mu$-like event distributions in the SK can be explained by the 3-flavor oscillations in Eq.~\eqref{eq:PoscModel}, taking into account the matter effect on $\nu_e$, which travel through the Earth. Indeed, 
for the electron neutrinos propagating in the matter with the electron number density $N_e$ the probability to oscillate into other flavor state can be written as~\cite{PDG2012,Wolfenstein:1977ue,Mikheev:1986gs,Barger:1980tf}
\begin{eqnarray}
	P_m(\nu_e\to\nu_{\alpha\neq e})	=	\sin^22\theta_m  \sin^2 \left( 	\frac{\Delta M^2 L}{ 4E}	\right),
\end{eqnarray}
where
\begin{eqnarray}\label{eq:sin2matt}
	\sin2\theta_m  = 	\frac{\tan2\theta}{\sqrt{\left(1 - \frac{N_e}{N_e^{\rm res}} \right)^2 + \tan^22\theta }},
\end{eqnarray}
\begin{eqnarray}
	\Delta M^2 	= 	 \Delta m^2  \left[	\left(1 - \frac{N_e}{N_e^{\rm res}} \right)^2 \cos^22\theta + \sin^22\theta	\right]^{\frac{1}{2}},
\end{eqnarray}
and the Mikheyev-Smirnov-Wolfenstein (MSW) resonance density is given by
\begin{eqnarray}\label{eq:res_density}
	N_e^{\rm res}	&=&	\frac{\Delta m^2 \cos2\theta}{2\sqrt{2}E G_\text{F}}  \\
	&\approx& 	6.56\times10^6 \,	\frac{\Delta m^2 [\text{eV}^2]}{E [\text{MeV}]}	\cos2\theta	\quad	\text{cm}^{-3} \, \text{N}_\text{A}  \nonumber
\end{eqnarray}
with $G_\text{F}$ and $\text{N}_\text{A}$ being Fermi constant and Avogadro number, respectively. 
In particular, for the atmospheric neutrinos with energies $E\sim10$~GeV we have $N_e^{\rm res} \approx 1.15~\text{cm}^{-3} \, \text{N}_\text{A}$, and using the mean electron number density in the Earth core $\bar N_e^c \approx 5.4~\text{cm}^{-3} \, \text{N}_\text{A}$~\cite{PDG2012,Dziewonski:1981xy}, we find 
\begin{eqnarray}
	P_m(\nu_e\to\nu_{\alpha\neq e})	=	0.05 \,  \sin^2 \left( 	2.8\frac{\Delta m^2 L}{ 4E}	\right),
\end{eqnarray}
which is significantly suppressed with respect to the vacuum oscillations in Eq.~\eqref{eq:PoscModel}. 
Notice that the $\nu_\mu \leftrightarrow \nu_\tau$	oscillations in the matter of the Earth proceed practically as in
vacuum due to approximate equality of the refraction indices~\cite{PDG2012}.

\subsection{$\nu_\mu$ and $\bar\nu_\mu$ oscillations} 

The close to unity amplitude of the muon neutrino oscillations $A_\mu$ observed in the MINOS experiment~\cite{Adamson:2013whj,Adamson:2011ig} is explained in DNT with tiny perturbations by the large oscillations of the muon neutrinos to both electron and tau neutrino flavors,  and analogous for the antineutrinos. The matter effect plays a subdominant role for the beam neutrinos in the MINOS experiment in contrast to the nearly upward atmospheric neutrino events 
due to the shorter baseline $L=734$~km and relatively small matter density in the Earth crust $\bar N_e \approx 2~\text{cm}^{-3} \, \text{N}_\text{A}$. Neglecting the matter effect we have the probability
\begin{eqnarray}\label{eq:PoscSurvival}
	P(\nu_\mu\to\nu_{\alpha\neq\mu})	&=&		P(\nu_\mu\to\nu_e)	 +	P(\nu_\mu\to\nu_\tau)	\nonumber\\
								&\approx&		\frac{8}{9}	\sin^2 \left( 	\frac{\Delta m^2 L}{ 4E}	\right),
\end{eqnarray}
whose amplitude agrees with the MINOS result $\nobreak{A_\mu>0.890}$ ($\nobreak{A_\mu>0.83}$) at 90\% CL 
for the muon (anti)neutrino oscillations, derived within 2-flavor approximation~\cite{Adamson:2013whj}. Reanalysis of the MINOS data taking into account the matter effect in the framework of 3-flavor democratic oscillations would even improve this agreement~\cite{1304.4870}.

We remark that in CNT $\nu_\mu\to\nu_\tau$ oscillations dominate in the muon neutrino disappearance in MINOS, and 
$A_\mu$ is considered as an amplitude $4|U_{\mu3}|^2|U_{\tau3}|^2$ of these $\nu_\mu\to\nu_\tau$ oscillations 
with close to maximal ($1/\sqrt{2}$) values of $|U_{\mu3}|$ and $|U_{\tau3}|$.

\subsection{Solar neutrinos} \label{Solar neutrinos}

\vspace{-8mm}
\begin{center}
 \begin{figure}[tb]
 \centering
	\includegraphics[width=0.47\textwidth]{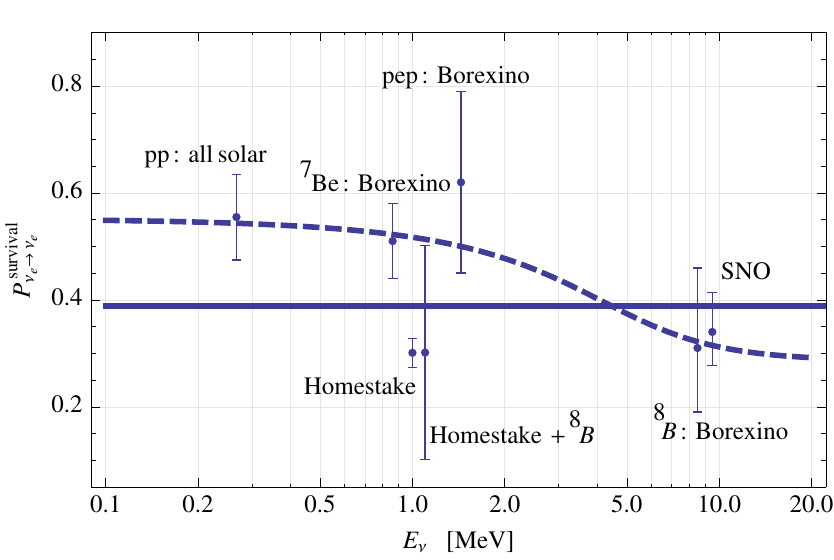}
   \caption{Survival probability for solar neutrinos $P_{ee}\equiv P_{\nu_e\to\nu_e}^\text{survival}$ vs. the neutrino energy. Solid (dashed) line represents the result of DNT with tiny perturbations (CNT).}
   \label{Fig:solar_nu}
 \end{figure}
\end{center}
Due to the combined effort of the fusion reactions, which produce the neutrinos in the core of the Sun
\begin{eqnarray}
	4p	\to	^4\text{He} + 2e^+ + 2\nu_e,
\end{eqnarray}
Sun is a source of the electron neutrinos.  
The observed flux of the solar neutrinos $\nu_\odot$ reveals a deficit with respect to the predictions of the standard solar model, which is known as the solar neutrino problem~\cite{PDG2012}. In other words, the survival probability $P_{\nu_e\to\nu_e}^\text{survival}$ for solar neutrinos is suppressed with respect to unity, which is shown in Fig.~\ref{Fig:solar_nu} by the data points obtained by the Homestake~\cite{Cleveland:1998nv}, the SNO~\cite{Aharmim:2005gt}, the Borexino~\cite{Bellini:2013lnn,Calaprice:2012kc}, and other solar neutrino experiments~\cite{Giunti:2007ry}. However part of the shown results for  $P_{\nu_e\to\nu_e}^\text{survival}$ were determined by comparing the measured fluxes with standard solar model predictions~\cite{Calaprice:2012kc}.

In DNT with tiny perturbations the term in $P_{\nu_e\to\nu_e}^\text{survival}$, which is proportional to $\cos(\Delta m^2L/2E)$, is suppressed due to the averaging over the region of $\nu_e$ production in the Sun, similarly to CNT~\cite{PDG2012}. On the other hand, the oscillations due to $\Delta\mu^2$ are strongly suppressed by the effect of solar matter if $\Delta\mu^2\ll10^{-7}$. 
Hence from Eq.~\eqref{eq:P_surviv} we get an incoherent result\footnote{For the interchanged first and third rows of $U_0$ in Eq.~\eqref{eq:U} we have $P_{\nu_e\to\nu_e}^\text{survival}	=	5/9$, which agrees only with a part of the $\nu_\odot$ data.} 
\begin{eqnarray}\label{eq:Psolar}
	P_{\nu_e\to\nu_e}^\text{survival}	=	\frac{7}{18},
\end{eqnarray}
which is represented by the solid line in Fig.~\ref{Fig:solar_nu}. This line comes to an agreement with the data in case of slightly less conservative error bars for the low energy data points. 
In fact, it is reasonable to reexamine these error bars since the low energy $\nu_\odot$ data still suffer from the Gallium anomaly~\cite{Giunti:2012tn,Abdurashitov:2005tb}, the discrepancy between the Homestake and the Borexino results, effects of solar model calculations, etc. One example of such reexamination was done in Ref.~\cite{Calaprice:2012kc}, where the Homestake data were combined with the Borexino $^8$B data (the result is also shown in Fig.~\ref{Fig:solar_nu}). 

We should stress that $P_{\nu_e\to\nu_e}^\text{survival}$ in Eq.~\eqref{eq:Psolar} is independent of the $\nu_\odot$ energy. This provides an efficient tool for verification of the solar model calculations, since the neutrino fluxes are quite sensitive to their parameter variations, in particular, in the solar models with late accretion~\cite{Serenelli:2011py}. 

The dashed line in Fig.~\ref{Fig:solar_nu} represents the result of CNT, where the transition from higher to lower values of $P_{\nu_e\to\nu_e}^\text{survival}$ with the energy increase is due to the MSW resonant neutrino oscillations in the Sun. This line can be also approximately reproduced in the DNT with sizable perturbations.\footnote{Note that the requirement to reproduce the decreasing of $P_{\nu_e\to\nu_e}^\text{survival}$ with the energy increase, having either a mixing matrix constructed of the eigenvectors of $M_0$ in Eq.~\eqref{eq:mass_matrices} or a slightly perturbed one, restricts possible form of the leading order mixing matrix. One of the possibilities is $U_0$ in Eq.~\eqref{eq:U} with zero ${\tau1}$ entry. Another allowed matrix with same third column as in $U_0$ has zero ${\mu1}$ entry. }

\section{Discussion}\label{Discussion}

There is a large number of (anti)neutrino flux calculations and Monte Carlo event generators in the market, 
and large number of discrepancies among them~\cite{Christensen:2013eza,Paukkunen:2013qfx,Mention:2011rk,Cao:2011gb}. These generators were developed using the experimental data, but in the framework of CNT. From the point of view of DNT, this is an additional source of the systematic errors. 
For this reason, so far we have discussed the neutrino experimental results, which are less dependent on the absolute normalization of the data samples.

\begin{table}[htdp]
\caption{Comparison of democratic and conventional models}
\vspace{-2mm}
\begin{center}
\begin{tabular}{|l|| l | l |}
	\hline
	 		 Experimental			& CNT 	& DNT with tiny \\
				 result		&  	& perturbations   \\
	\hline
	\hline
		$\nu_e\to\nu_{\mu,\tau}$   &  Small $|U_{e3}|^2$	& Earth matter effect  \\
		suppression	&	(2-flavor oscillations)				& (3-flavor oscillations)	\\
	\hline
		Large $A_{\nu_\mu}$ & Large $|U_{\mu3}|^2$ and $|U_{\tau3}|^2$ 	& Cumulative effect of \\
			& (about 1/2 each) in the	& $\nu_\mu\to\nu_\tau$ and $\nu_\mu\to\nu_e$	\\
						& probability of $\nu_\mu\to\nu_\tau$ &	 oscillations	\\
	\hline
		$\nu_\odot$ deficit 	& MSW resonant  			& Averaged \\
						& oscillations 		& $P_{\nu_e\to\nu_e}^\text{survival}	=	7/18$	\\
	\hline
\end{tabular}
\end{center}
\label{Tab:1}
\end{table}%
 The differences in the explanation of the discussed neutrino data in the framework of CNT and DNT are summarized in Table~\ref{Tab:1}. Its left column lists the following experimental results: 
 (1) suppression of $\nu_e\to\nu_{\mu,\tau}$ oscillations observed by the SK, (2) large amplitude of muon neutrino oscillations observed by the MINOS and the SK, and (3) deficit of solar neutrinos. Ways of explanation of these results within CNT (which can be applied also in the DMT with sizable perturbations) are listed in the second column. The third column shows the alternative ways to explain these results within the DNT with tiny perturbations. Note that in this case only the parameter $m$ significantly effects the discussed observables. This makes it possible to perform a simplified global analysis of the neutrino data. However the experiments, which measure $\theta_{13}$ should be properly taken into the consideration.

\subsection{Measurements of $\theta_{13}$}

Several experiments~\cite{Adamson:2011qu,Abe:2011fz,An:2012eh,Ahn:2012nd} have interpreted their data as measuring of $\theta_{13}\approx15^\circ$.
However these results depend significantly on the discussed above normalization of the neutrino fluxes, the assumption of $\sin^22\theta_{23}=1$, etc. 
For example, in the reactor experiments~\cite{Abe:2011fz,An:2012eh,Ahn:2012nd} $\bar\nu_e$ disappearance ratio is
\begin{eqnarray}
	R \equiv \frac{N^\text{obs}}{N^\text{theor}} = 1 - \sin^22\theta_{13} \sin^2 \left( \frac{1.27\Delta m^2 L[m]}{E_{\bar\nu_e}[MeV]} \right),
\end{eqnarray}
where $N^\text{obs}$ ($N^\text{theor}$) is the observed (expected, assuming no oscillations) number of the antineutrinos. 
As a result, in case of close to unity value of $R$ (small $\theta_{13}$), which is supposed in CNT, a 10\% error in the total neutrino flux leads to a huge error in $\theta_{13}$ exceeding 100\%. On the other hand, in case of smaller $R$ the flux calculations derived within CNT (including the correlations among reactor cores, etc.) can not be applied in DNT. 
We argue that the measurements of $\theta_{13}$ should be reconsidered from the stage of development of the event generators in the framework of DNT, in order to seriously verify this theory.


\section{Predictions}\label{Predictions}

In DNT with tiny perturbations the neutrino masses and mixings are determined with good precision using the atmospheric neutrino data. This allows to make interesting predictions discussed below.

\subsection{Direct neutrino mass experiments} 

Using Eqs.~\eqref{eq:spectrum_perturbed} and \eqref{eq:Uperturbed}, the average mass, determined through the analysis of low energy beta decays, can be written as
\begin{eqnarray}\label{eq:m_beta}
	\langle m_\beta \rangle 	\equiv	\sqrt{ \sum_i m_i^2 |U_{ei}|^2 }	 =	m\sqrt{2} +	{\cal O}(m^{3/4}\delta m^{1/4}).
\end{eqnarray}
For $m\approx0.03$~eV from Eq.~\eqref{eq:nu_mass_scale} we have $\langle m_\beta \rangle \approx0.04$~eV, which is far below the present upper limit $\langle m_\beta \rangle < 2.5$~eV (at 95\% CL)  obtained in the ``Troitsk $\nu$-mass" experiment~\cite{Lobashev:2001uu}, and below the sensitivity $0.2$~eV (at 90\% CL) of ongoing {\nobreak KATRIN} experiment~\cite{Titov:2004pk,Formaggio:2012zz}.  However the new approaches, such as planned MARE, ECHO, and Project8 experiments, may probe the neutrino mass in the sub-eV region~\cite{Drexlin:2013lha}.

\subsection{Neutrinoless double beta decay}

Using Eqs.~\eqref{eq:spectrum_perturbed} and \eqref{eq:Uperturbed}, the effective Majorana mass in the neutrinoless double beta decay can be written as
\begin{eqnarray}\label{eq:m_beta-beta}
	\langle m\rangle 	\equiv	\sum_i s_i m_iU_{ei}^2	=	{\cal O}\left(\sqrt{m\delta m}\right).		
\end{eqnarray}
where the terms of order $m$ cancel each other due to the sign factors in Eq.~\eqref{eq:Lmass}. Hence this decay can not be observed in the near future searches, which makes them a useful tool for studying either precision nuclear physics or some new physics~\cite{Ali:2010zza,Ali:2007zza,Ali:2007ec}.

Note that in Eqs.~\eqref{eq:m_beta} and \eqref{eq:m_beta-beta} we took into account possible deviations of order $(\delta m/m)^{1/2}$ in the elements of the first row of the matrix $U$ in Eq.~\eqref{eq:Uperturbed}.

\subsection{Neutrino magnetic moment}

The Majorana neutrinos do not have diagonal magnetic moments, while their transition magnetic moments in case of opposite $CP$ phases ($CP$ parities) are nonzero~\cite{Broggini:2012df}: 
\begin{eqnarray}
		\mu_{ij}^\text{Majorana} &=& 2\mu_{ij}^\text{Dirac}
			\approx	-8\times10^{-24} ~\left( \frac{m_i + m _j}{0.1~\text{eV}} \right) 	\nonumber\\
			&\times&	\sum_{\ell=e,\mu,\tau} \left(\frac{m_\ell}{m_\tau}\right)^2 U_{\ell i}^*U_{\ell j}	~\mu_B,  \label{eq:NMMlimitSMtransit}
\end{eqnarray}
where $\mu_B=e/(2m_e)=5.788 \times10^{-5}$ eV\,T$^{-1}$ is the Bohr magneton, and $m_\ell$ are the charged lepton masses. 

	Using Eqs.~\eqref{eq:U}, \eqref{eq:spectrum} and \eqref{eq:nu_mass_scale}, 
we have  $\mu_{23}\approx 3.4\times 10^{-24}~\mu_B \gg \mu_{12}, \mu_{13}$. The effects of perturbations can not increase this tiny scale of the neutrino magnetic moment, 
which is 13 orders of magnitude below the present terrestrial bound $\mu_{\bar\nu_e}\approx 2.9\times 10^{-11}~\mu_B$ (90\% CL)~\cite{Beda:2012zz}. 
This leaves a good opportunity for the new physics searches~\cite{Zhuridov:2013ika,Bellazzini:2010gn}.

\section{Conclusion} 

A neutrino theory, which is based on the simple democratic neutrino mass matrix, is introduced. This theory predicts normal hierarchical neutrino mass spectrum with the absolute neutrino mass scale of $m\approx 0.03$~eV. 
In the case of tiny perturbations to the democratic mass matrix this theory explains to a good accuracy the atmospheric, solar and accelerator neutrino data, using only the parameter $m$, and has a significant predictive power. Interestingly, in this case we use an alternative explanation of several important neutrino results. Nevertheless, in the case of sizable perturbations the conventional explanation may be applied. 

%

\section*{Acknowledgements} 

This work was supported in part by the US Department of Energy under the contract DE-SC0007983. The author  
thanks Lincoln Wolfenstein and Alexey Petrov for useful discussions. The author also thanks Boris Altshuler, Anatoly Borisov, Gil Paz and Matthew Gonderinger for useful discussions and comments on the manuscript.

\end{document}